\documentclass[a4paper,12pt]{article}
\usepackage{amsmath,amssymb,amsthm,mathtools}

\usepackage[colorlinks,citecolor=blue,urlcolor=magenta]{hyperref}
\usepackage{doi}
\usepackage{booktabs}
\usepackage[]{csquotes}
\usepackage[american]{babel}
\usepackage[letterpaper,margin = 1.25in]{geometry}
\usepackage[style=
authoryear-comp, 
sorting=nyt, 
dashed=false, 
maxcitenames=2, 
maxbibnames=99, 
uniquelist=false,
uniquename=false,
giveninits=true, 
natbib, 
date=year 
]{biblatex}

\AtBeginRefsection{\GenRefcontextData{sorting=ynt}}
\AtEveryCite{\localrefcontext[sorting=ynt]}

\DeclareFieldFormat{pages}{#1} 
\renewbibmacro{in:}{\ifentrytype{article}{}{\printtext{\bibstring{in}\intitlepunct}}} 

\usepackage[inline,shortlabels]{enumitem}
\setlist[enumerate,1]{label=(\roman*)}

\usepackage{setspace}
\title{Proof of Steady-State Multiplicity in Aiyagari\thanks{I thank Dan Cao, Johann Fuchs, Elisabeth Proehl, and Eric Young, discussions with whom inspired me to pursue this work, and I thank Lint Barrage and Alexis Akira Toda for helpful comments.}
}

\author{Kieran James Walsh\thanks{ETH Zurich, KOF Swiss Economic Institute. Email: \href{mailto:kwalsh@ethz.ch}{kwalsh@ethz.ch}.}}

\date{\today}

\theoremstyle{plain}
\newtheorem{thm}{\bf Theorem}
\newtheorem{lem}[thm]{\bf Lemma}
\newtheorem{cor}[thm]{\bf Corollary}
\newtheorem{prop}[thm]{\bf Proposition}

\numberwithin{equation}{section}
\numberwithin{thm}{section}

\begin{document}
\maketitle

\begin{abstract}
 I provide the first analytic construction of a canonical \citet{Aiyagari1994} economy exhibiting at least three steady states. Along the way, I provide new upper and lower bounds on the stationary capital supply for the case where the net return on saving is negative. I also give parameter restrictions that guarantee the existence of a steady state with a rental rate in the neighborhood of a remarkably simple number: the capital share times the depreciation rate ($\alpha \times \delta$).

\medskip


\medskip

\end{abstract}

\section{Introduction}\label{sec:intro}

The model of \citet{Aiyagari1994} is one of the cornerstones of macroeconomics.\footnote{Along with \citet{Imrohoroglu1989} and \citet{Huggett1993}, \citet{Aiyagari1994} is a seminal work of the broader ``Bewley'' (e.g., \citet{Bewley1986}) class of models.} In the model, infinitely-lived agents facing idiosyncratic income risk and borrowing constraints smooth consumption by saving capital, which is rented to a representative firm that also hires the agents as workers. There are no aggregate shocks. In equilibrium, rental rates and wages are such that firm capital demand is equal to the total supply of capital from the optimizing agents. Although the model's equilibrium has been widely explored with computational approximations, its theoretical properties have remained relatively elusive. For example, most existing theoretical results pertain to steady states (``stationary equilibria''), little is known in practice about stability and determinacy, no papers theoretically study equilibrium far away from steady states, and, to my knowledge, no papers investigate the potential for non-stationary equilibria. And while strong conditions for steady-state uniqueness have been established (see the survey of \citet{TodaWalsh2024}), and two papers offer computational examples seeming to exhibit a multiplicity of steady states \citep{Acikgoz2018,WalshYoung2026}, the theoretical possibility of multiplicity has never been proven.

In this paper, I provide an example in which I can prove that the \citet{Aiyagari1994} model has at least three steady states (stationary equilibria). The example's parameters are not a standard calibration of the model, but the setup itself is the canonical one. My approach consists of creating bounds on the stationary supply of capital, building on the work of \citet{Acikgoz2018}. \citet{Acikgoz2018} proves existence of at least one steady state, via the intermediate value theorem, by establishing: (1) capital demand exceeds the stationary capital supply at a sufficiently low rental rate (call it $\underline{r}$), (2) capital supply exceeds demand at a sufficiently high rental rate ($\bar{r}$), and (3) supply and demand are continuous in the rental rate. It follows that supply and demand must intersect at some intermediate rate. 

To establish multiplicity, I develop new upper and lower bounds for stationary capital supply that apply when the return on saving is negative and there are two individual income states.\footnote{Having only two states simplifies the analysis, but there is no reason to think two is special and, intuitively, more states and thus more degrees of freedom should broaden the scope for multiplicity.} For my example set of parameters, I analyze the new bounds at two intermediate rental rates, $r_L$ and $r_H$. Letting $\delta \in (0,1)$ denote the depreciation rate and $\beta \in (0,1)$ denote the discount factor, the rates satisfy $0 < \underline{r} < r_L < r_H < \delta < \bar{r} < 1/\beta -1 + \delta$. I show that the \emph{lower} bound on capital supply exceeds capital demand at $r_L$, and the \emph{upper} bound on capital supply is less than capital demand at $r_H$. Combining this with \citet{Acikgoz2018}'s bounds and continuity result, there must be at least 3 steady states in my example (again, by the intermediate value theorem). 

The steady state rental rate between $\underline{r}$ and $r_L$, which I will call $r_{\alpha \delta}$, has a surprising property. \citet{WalshYoung2026}, who construct hundreds of \citet{Aiyagari1994} multiplicity examples via numerical approximation, called $r_{\alpha \delta}$ the ``very low rate'' steady state. They found that $r_{\alpha \delta}$ was usually close to the number $\alpha \times \delta$, and they offered an intuitive but not fully-rigorous explanation for this phenomenon. Here, in Theorem \ref{thm:verylow}, I validate their intuition and prove that under a simple parameter restriction there always exists a steady state in the neighborhood of $\alpha \times \delta$, given that the worst case scenario income realization is taken to be sufficiently small. As we will see, Theorem \ref{thm:verylow} is important because it facilitates the construction of multiplicity examples.

The key lower bound on supply in my analysis (Equation \ref{eq:LB}) is based on the observation of \citet{Acikgoz2018} and \citet{WalshYoung2026} that, when there is a disaster individual income state and the return on saving is low, agent consumption/saving decisions are driven by worry about the disaster state. The key upper bound I establish (Proposition \ref{prop:UB}) is more technical and relies on the \citet{coleman1990solving} operator approach of \citet{Acikgoz2018} and \citet{li2014solving}. The new bounds I construct apply when the return on saving is negative: $1 + r - \delta <1$ or, equivalently, $r<\delta$. I conjecture that negative returns are necessary for multiplicity in the \citet{Aiyagari1994} model, but I do not attempt to prove that here. 

Beyond \citet{WalshYoung2026}, \citet{Acikgoz2018}, and \citet{Acikgoz2018}'s precursor \citet{li2014solving}, this paper is related to the broader literature on uniqueness/multiplicity in Bewley models reviewed in \citet{TodaWalsh2024}. \citet{AchdouHanLasryLionsMoll2022}, \citet{Toda2019JME}, and \citet{Light2020,Light2023} all highlight that relative risk aversion less than one implies steady-state uniqueness in Bewley models. \citet{Proehl2018} provides a uniqueness result for recursive competitive equilibrium, which, as described in \citet{WalshYoung2026}, potentially applies to \citet{Aiyagari1994} steady states. Probably the only analytic multiplicity example in the literature is in \citet{Toda2017JEDC}, in the context of the \citet{Huggett1993} model without a borrowing constraint and with CARA utility. I contribute to this literature by definitively proving multiplicity is possible in the canonical \citet{Aiyagari1994} model and by offering new bounds on saving, bounds which apply when the return on saving is negative.

Section \ref{sec:model} describes the \citet{Aiyagari1994} model, Section \ref{sec:LB} establishes the lower bound and very low rate steady state result, Section \ref{sec:UB} uses the \citet{coleman1990solving} operator approach to get the upper bound, Section \ref{sec:mult} gives the multiplicity example, Section \ref{sec:conclusion} concludes, and all proofs are in Appendix \ref{app:proofs}.

\section{Model}\label{sec:model}

There are an infinite number of discrete time periods $t=0,1,2,\dots$ and a continuum of agents of measure one (indexed by $i$). The agents are ex ante identical but have idiosyncratic labor supply processes. An agent's exogenous labor supply ($e$) follows a Markov process with $S$ states $e_1<e_2<\dots<e_{S}$ and irreducible transition matrix $P$ satisfying $P(e_1,e_1)>0$.\footnote{The irreducibility of $P$ and persistence of the lowest state constitute Assumption 4 in \citet{Acikgoz2018}.} Let $\mathcal{E}$ denote the set of labor supply states. $e_1 = \varepsilon>0$ is the lowest labor supply realization, and $e_S=H$ is the highest. Let $p_s$ denote the stationary probability of state $s \in S$, and let $L = \sum_{s = 1}^S p_ie_i$ denote the stationary aggregate supply of labor.

For my bounds and multiplicity results in subsequent sections, I will impose the two-state version of the model with $S=2$, although similar arguments should apply to the $S>2$ case. Indeed, the computational examples in \citet{WalshYoung2026} and \citet{Acikgoz2018} have $S>2$. Here in this section, I maintain arbitrary $S < \infty$ to highlight existing knowledge from the literature.

Given initial individual capital $a_0 \geq 0$, an agent chooses paths for consumption $c_t$ and capital saving $a_{t+1}$ to solve
\begin{align}\label{eq:agentprob}
    & \max_{c_t \geq 0,a_{t+1}  \geq 0}  E\sum_{t=0}^\infty \beta^t \frac{c_t^{1-\gamma}}{1-\gamma} \ \text{subject to} \\
    & c_t + a_{t+1} \leq w e_t + (1+r - \delta)a_t,\nonumber
\end{align}
where $\beta \in (0,1)$ is the discount factor, $\gamma>1$ is the coefficient of relative risk aversion (flow utility is of the CRRA form), and $\delta \in (0,1]$ is the rate of depreciation of individual capital.\footnote{I maintain $\gamma>1$ throughout because it is already known that $\gamma \leq 1$ is sufficient for steady-state uniqueness.} $w$ is the wage, $r$ is the rental rate, and I assume $w >0$ and $0 < r < 1/\beta - 1 + \delta$. This implies that the gross return on saving $R = 1 + r - \delta$ satisfies $1-\delta < R < 1/\beta$. When $R<1$ or, equivalently, $r<\delta$, the return on saving is negative. We can rewrite the budget constraint in \eqref{eq:agentprob} as
\begin{equation}\label{eq:budget}
    c_t + a_{t+1} \leq w e_t + Ra_t.
\end{equation}
Note that borrowing is not allowed ($a_{t+1} \geq 0$).

A representative firm hires labor and rents capital to solve
\begin{equation*}
    \max_{k \geq 0,\ell \geq 0} k^\alpha \ell^{1-\alpha} - rk - w\ell,
\end{equation*}
where $\alpha \in (0,1)$ is the capital share. Firm optimization requires
\begin{align}
    k & = K(r,\ell) \equiv \left( \frac{\alpha}{r} \right)^{\frac{1}{1-\alpha}} \ell \label{eq:firm1} \\
    w & = (1-\alpha) \left( \frac{\alpha}{r} \right)^{\frac{\alpha}{1-\alpha}}. \label{eq:firm2}
\end{align}
This setup is the canonical \citet{Aiyagari1994} model, and it satisfies Assumptions 1--5 of \citet{Acikgoz2018}, which he uses to establish all of his results I use below. 

By \citet{Acikgoz2018} and \citet{li2014solving}, the agent problem \eqref{eq:agentprob} is solved by the unique fixed-point consumption function $c(a,e)$ of a \citet{coleman1990solving} operator $\mathcal{K}$. $\mathcal{K}$ maps $\mathcal{C}$ into itself, where $\mathcal{C}$ is the appropriately-defined consumption function space,\footnote{\label{fn:coleman}Following \citet{Acikgoz2018}, $\mathcal{C}$ is the set of functions $c(a,e):\mathbb{R}_+ \times \mathcal{E} \rightarrow \mathbb{R}_+$ such that, for all $(a,e) \in \mathbb{R}_+ \times \mathcal{E}$,  (i) $c$ is continuous, (ii) $c$ is weakly increasing in $a$, and (iii) $0<c(a,e) \leq we + Ra$, and $\sup \left| c(a,e)^{-\gamma} - (we + Ra)^{-\gamma} \right| < \infty$.} and $\mathcal{K}$ is implicitly given by
\begin{equation}\label{eq:Coleman}
    \left(\mathcal{K}c(a,e)\right)^{-\gamma} = \max \left\{ \beta R E_e \left[ c\left( we + Ra - \mathcal{K}c(a,e)    , e' \right)^{-\gamma}   \right]  , \left(we +  Ra \right)^{-\gamma}  \right\}.
\end{equation}
Paths generated by the fixed point $\mathcal{K}c = c$ satisfy the standard Euler equation $c_t^{-\gamma} \geq \beta R E_t\left[ c_{t+1}^{-\gamma} \right]  $, and, as will be important for deriving the upper bound below, $\mathcal{K}$ is monotone.

Next, define the optimal saving function $a' = g(a,e) \equiv we + Ra - c(a,e)$. \citet{Acikgoz2018} shows that, for any $w>0$ and $R \in (1-\delta,1/\beta)$, $g$ generates a unique stationary cross-sectional distribution $\Omega(R,w)$ over capital and labor $(a,e)$: initializing the economy at $\Omega$, the optimal saving policies yield $\Omega$ in the next period also. $A(R,w) \equiv \int g(a,e) d\Omega(R,w)$ is then the stationary aggregate supply of capital.

A steady state, or stationary equilibrium, consists of prices $(R,w)$ such that capital and labor markets clear at the stationary cross-sectional distribution when agents and firms are optimizing: $A(R,w) = K(R-1 + \delta,L)$ and $w  = (1-\alpha) \left( \frac{\alpha}{R-1+\delta} \right)^{\frac{\alpha}{1-\alpha}}$. In other words, \eqref{eq:firm2} holds and optimal stationary capital supply ($A$) and stationary labor supply ($L$) satisfy firm demand.\footnote{Technically, a stationary equilibrium is a steady state of the more general rational expectations equilibrium, which has potentially time-varying prices and cross-sectional distributions. See \citet{WalshYoung2026} for a discussion of this point.}

We can simplify the search for steady states by further exploiting the CRRA and Cobb-Douglas forms. As \citet{Acikgoz2018} shows, CRRA implies that the stationary capital supply is homogeneous of degree one in the wage. Using the firm FOC \eqref{eq:firm2}, we can rewrite the steady state condition $A(R,w) = K(R-1 + \delta,L)$ as
$$ A(R,1) = \frac{K(R-1 + \delta,L)}{ (1-\alpha) \left( \frac{\alpha}{R-1+\delta} \right)^{\frac{\alpha}{1-\alpha}}}, $$ and the RHS simplifies to $$ D(R) \equiv  \frac{\alpha}{1-\alpha} \frac{L}{R -1 + \delta}.  $$
That is, it is without loss of generality to study the model with the wage normalized to one and firm capital demand appropriately adjusted.

Henceforth, I normalize the wage to one, write $A(R)$ instead of $A(R,1)$, and call $D(R)$ capital demand (even though it is really firm capital demand divided by the wage expression \eqref{eq:firm2}). Defining $Z(R) = A(R) - D(R)$ to be excess stationary capital supply in the adjusted economy, steady states of the true economy are given by $Z(R) = 0$.

\section{Lower Bound and the Very Low Rate Steady State}\label{sec:LB}

Now I specialize to $S=2$, implying that $\mathcal{E}= \{\varepsilon, H\}$. Let $p_\varepsilon$ and $p_H$ denote the stationary labor supply probabilities. Let $\pi = P(\varepsilon,H) $ be the probability of transitioning from labor supply state $\varepsilon$ to $H$, and let $\lambda$ be the $H$ to $\varepsilon$ probability. As I will assume $\varepsilon$ is a very small number later, $\lambda$ is the probability of entering an individual income disaster. Low $\pi$ and $\lambda$ indicate high persistence of individual labor income, and we have that $p_\varepsilon = \lambda/(\lambda + \pi) $ and $p_H = \pi/(\lambda + \pi)$, by standard Markov process properties.

My new lower bound on capital supply $A$ is based on the idea that, for sufficiently low return rates on saving, agents will precautionarily accumulate capital to protect against long sequences of terrible income realizations. When $\varepsilon$ is close to zero, concern about such sequences creates a tight lower bound on aggregate saving. In this section, I assume $R< 1$, or, equivalently, $r<\delta$ (the return on saving is negative).

The starting point for the bound is the Euler equation $c_t^{-\gamma} \geq \beta R E_t\left[ c_{t+1}^{-\gamma} \right]$. Now fix $N>1$. Iterating forward the Euler equation to period $t+N$, it becomes $c_t^{-\gamma} \geq (\beta R)^N E_t\left[ c_{t+N}^{-\gamma} \right]$.

Given $(a,e)$ in time $t$, consider the $N$-period sequence of income realizations $e_{t+1} = \varepsilon,\dots,e_{t+N} = \varepsilon$, which is the worst case scenario over the next $N$ periods. For $e_t = \varepsilon$ and $e_t = H$ respectively, the probabilities of these paths are $q_N(\varepsilon)  = (1-\pi)^N$ and $q_N(H) = \lambda(1-\pi)^{N-1}$. Let $c_N(a,e)$ denote consumption in period $t+N$ conditional on time $t$ state $(a,e)$ and the $N$-period worst case scenario path. Since the worst case scenario is only one branch in the expectation $E_t$, we must have
\begin{align}
    c(a,e)^{-\gamma} \geq & (\beta R)^Nq_N(e) c_N(a,e)^{-\gamma} \implies \nonumber \\
      c_N(a,e) \geq & \left( (\beta R)^Nq_N \right)^{1/\gamma} c(a,e), \label{eq:MULB}
\end{align}
where $q_N = \min\{q_N(\varepsilon),q_N(H)\}$.

Next, we can put an upper bound on $c_N(a,e)$. The most that could be consumed at $t+N$ along the worst case scenario path would come from saving the maximal feasible amount at $t+1,\dots,t+N-1$ and then consuming the most possible at $t+N$. Recalling that $g(a,e) = e + Ra - c(a,e)$ is this optimal saving policy at $t$, this gives
\begin{align}
c_N(a,e) & \leq R^N g(a,e)  + R^{N-1}\varepsilon + \dots + R\varepsilon + \varepsilon  \nonumber \\
 & =  R^Ng(a,e) + \frac{1-R^N}{1-R} \varepsilon.  \label{eq:cnUB}
\end{align}

Combining \eqref{eq:MULB} and \eqref{eq:cnUB} and rearranging, we arrive at
\begin{equation*}
     R^Ng(a,e) \geq \left( (\beta R)^Nq_N \right)^{1/\gamma} c(a,e) - \frac{1-R^N}{1-R} \varepsilon.
\end{equation*}
Integrating this inequality with respect to the stationary distribution $\Omega(R)$ implies
\begin{equation}\label{eq:AR1}
    R^N A(R) \geq \left( (\beta R)^Nq_N \right)^{1/\gamma} C(R) - \frac{1-R^N}{1-R} \varepsilon,
\end{equation}
where $C(R)$ is stationary aggregate consumption. Integrating the budget constraint \eqref{eq:budget}, which always binds by non-satiation, with respect to the stationary distribution implies $C(R) = L -  (1-R)A(R)$. Plugging this into \eqref{eq:AR1} and rearranging, we have a new lower bound for the stationary supply of capital at arbitrary $R<1$:
\begin{equation}\label{eq:LB}
    A(R) \geq  \underline{A}_{N,\varepsilon}(R) \equiv \frac{\left( (\beta R)^Nq_N \right)^{1/\gamma} L_\varepsilon - \frac{1-R^N}{1-R} \varepsilon}{R^N + (1-R)\left( (\beta R)^Nq_N \right)^{1/\gamma} },
\end{equation}
where I now write $L_\epsilon = p_\varepsilon \varepsilon + p_H H$ (instead of $L$) to note its dependence on $\varepsilon$.\footnote{One could alternatively normalize $L$ to some number, $1$ say, and have $H$ depend on $\varepsilon$, but my analysis would be the same.}

There is also a quite loose but easy-to-establish upper bound on capital supply that applies when $R<1$. I refer to this upper bound as $\bar{\bar{A}}(R)$, with double bars to distinguish it from the tighter upper bound I will derive in the next section. Integrating the budget constraint, using $C(R)\geq 0$, and rearranging, we get
\begin{equation}\label{eq:UB1}
    A(R) \leq \bar{\bar{A}}(R) \equiv \frac{L_\varepsilon}{1-R},
\end{equation}
so for $R<1$ we have $\underline{A}_{N,\varepsilon}(R) \leq A(R) \leq \bar{\bar{A}}(R)$. It is easy to check that $\underline{A}_{N,\varepsilon}(R) < \bar{\bar{A}}(R)$ by dividing the numerator and denominator of \eqref{eq:LB} by $\left( (\beta R)^Nq_N \right)^{1/\gamma}$.

The following proposition sets up the very low rate steady state result and the construction of the rental rates $0<\underline{r}<r_{\alpha \delta}<r_L$ described in the introduction. Later, we will have $Z(1+\underline{r}-\delta)<0$ and $Z(1+r_L-\delta)>0$, implying a steady state $r_{\alpha \delta} \approx \alpha \times \delta$.
\begin{prop}\label{lem:Alim}
    Fix $R <\left(\beta(1-\pi)\right)^\frac{1}{\gamma-1}$  and choose $\nu>0$. There exist $N^*>1$ and $\bar{\varepsilon}>0$ such that for all $\varepsilon \in (0,\bar{\varepsilon})$, $ \frac{L_\varepsilon}{1-R} - \underline{A}_{N^*,\varepsilon}(R)  <\nu$ and $ \frac{L_\varepsilon}{1-R} - A(R)  <\nu$.
\end{prop}

This proposition, which is new to the literature, establishes that when the return on saving is sufficiently low, both the lower bound on stationary capital supply and \emph{stationary capital supply itself} can be made arbitrarily close to $L_\varepsilon/(1-R)=L_\varepsilon/(\delta-r)$  (the loose upper bound $\bar{\bar{A}}$) by taking the disaster income state $\varepsilon>0$ sufficiently close to zero.\footnote{Note that $N^*$'s value is irrelevant here for the economics since it is a proof device and not a parameter affecting agent decisions per se. The key intuition is just that when both $R$ and $\varepsilon$ are very low, agents will be obsessed with the possibility of unlikely but disastrous draws of low income sequences.} Since $L_\varepsilon/(\delta-r)$ is equal to firm capital demand at $r=\alpha \times \delta$, this implies a steady state $r_{\alpha\delta} \approx \alpha \times \delta$ ($R_{\alpha\delta} \approx 1 - (1-\alpha)\delta$) when the disaster state is particularly disastrous (and other parameter restrictions hold). The formal result, which confirms the computational findings and intuition put forth in \citet{WalshYoung2026}, is given by the following theorem.
\begin{thm}\label{thm:verylow}
    Choose $\iota>0$. If $\left(\beta(1-\pi)\right)^\frac{1}{\gamma-1} > 1 - (1-\alpha)\delta $, then there exists a worst-case income $\bar{\varepsilon} > 0$ threshold such that, for any $\varepsilon \in (0,\bar{\varepsilon})$, there is a stationary equilibrium at some $r_{\alpha\delta}$ satisfying $\left| r_{\alpha \delta}-\alpha\delta \right| < \iota$.
\end{thm}
In short, if $\left(\beta(1-\pi)\right)^\frac{1}{\gamma-1}$ lies to the right of the gross return equivalent of $\alpha \times \delta$, then we can force a steady state arbitrarily close to $\alpha \times \delta$ by taking the worst-case income to be extremely small. So by searching in this parameter space, multiplicity just requires finding another steady state at a higher rate $r>\alpha \times \delta$, since we already know the very low rate steady exists. I do this below by finding a rental rate $r_H$ where the upper bound on capital supply is less than capital demand. This guarantees another steady because capital supply eventually diverges as the rate of return approaches the discount rate.

The fact that there will be at least \emph{three} steady states stems from the following corollary established in the proof of Theorem \ref{thm:verylow}.

\begin{cor}\label{cor:verylow}
    If $\left(\beta(1-\pi)\right)^\frac{1}{\gamma-1} > 1 - (1-\alpha)\delta $, then there exist rental rates $\underline{r}$ and $r_L$ and a worst-case income $\bar{\varepsilon} > 0$ such that, for any $\varepsilon \in (0,\bar{\varepsilon})$, there exists $r_{\alpha \delta}$ satisfying $\underline{r}<r_{\alpha \delta}<r_L$ and $Z(1+r_{\alpha \delta} - \delta) = 0$. Furthermore, the rental rates and excess supplies satisfy $0<\underline{r}<r_{\alpha \delta}<r_L < \delta$ and  $Z(1+\underline{r} - \delta) < Z(1+r_{\alpha \delta} - \delta) = 0 < Z(1+r_L - \delta) $.
\end{cor}

So when parameters are such that the very low rate steady state must exist, there is excess demand for capital at some rate below $r_{\alpha \delta}$ and excess supply of capital at some rate above $r_{\alpha \delta}$. If we can find $r_H \in (r_L,1/\beta - 1 +\delta)$ at which demand exceeds supply, there must be a third, intermediate steady state since supply exceeds demand at $r_L$. The next section shows how to find $r_H$. Of course, the very low steady state may be the only one, depending on the parameters, but knowing it is there simplifies the problem.

\section{Upper Bound}\label{sec:UB}

My new upper bound builds on the \citet{coleman1990solving} operator approach developed in \citet{Acikgoz2018} and \citet{li2014solving}. Indeed, my proof, which is technical but straightforward, closely follows \citet{Acikgoz2018}'s idea to build a ``subsolution'' to the implicit \citet{coleman1990solving} operator \eqref{eq:Coleman}. I simply guess a more general form of the subsolution that allows for a tight upper bound and accommodates a negative return on saving ($r<\delta$).\footnote{I initially attempted to use \citet{Acikgoz2018}'s upper bound, which only applies when $r>\delta$, but I was unable to find a multiplicity example after a long parameter search.}

Consider a (generally non-optimal) consumption rule of the form
\begin{equation}\label{eq:h}
    h(a,e) = \eta_e Ra + \omega_e e,
\end{equation}
where $\eta_e,\omega_e \in (0,1)$. This function is clearly in the feasible set defined by the budget constraint \eqref{eq:budget}, borrowing constraint $a'\geq0$, and $c\geq0$. Moreover, it is in the consumption function space $\mathcal{C}$, formally given in Footnote \ref{fn:coleman}, for the \citet{coleman1990solving} operator I called $\mathcal{K}$ \eqref{eq:Coleman}.

Suppose we had $ h \leq \mathcal{K}h$ for all $(a,e)$. Then, since $\mathcal{K}$ is monotone with unique fixed point $c$, we would have
$$ h \leq \mathcal{K}h \leq \mathcal{K}^2 h \leq \dots \leq c.    $$
That is, $h$ would be a subsolution to $\mathcal{K}$. This lower bound on the consumption function would then imply an upper bound on the saving function. This is exactly \citet{Acikgoz2018}'s approach to constructing an upper bound on saving, except he uses a form more restricted than \eqref{eq:h}, and his ultimate upper bound only applies when $R>1$. My general affine $h$ form allows for a tighter upper bound that applies when $R<1$. The main difficulty in my below argument is the phrase ``for all $(a,e)$,'' since $a \in \mathbb{R}_+$ in the consumption function space. But the following lemma implies that, since  we are analyzing stationary distributions, it is without loss of generality to take $a$ to have a finite bound in $\mathcal{C}$ when $R<1$.
\begin{lem}\label{lem:aub}
    Suppose $R \in (0,1)$. Then the budget constraint \eqref{eq:budget} and $c\geq 0$ imply $a' = g(a,e) \leq \bar{a} \equiv \frac{H}{1-R}   $ for any $(a,e) \in [0,\bar{a}] \times \mathcal{E}$.
\end{lem}
So the budget constraint and $c\geq0$ alone put an upper bound on $a_t$ when $R<1$ (this result was also in the proofs of \citet{Acikgoz2018}). Since $a'< a$ when $a>\bar{a}$, it is straightforward to show that the stationary distribution will never have mass on $a > \bar{a} $, and it is thus without loss of generality to bound $a$ in the function space $\mathcal{C}$.

Finally, the new upper bound on stationary capital supply is given in the following proposition (which also bounds individual saving more tightly than in Lemma \ref{lem:aub}):
\begin{prop}\label{prop:UB}
Suppose $0<R<1$, and let $h(a,e) = \eta_e Ra + \omega_e e$, with $\{\eta_\varepsilon,\omega_\varepsilon,\eta_H,\omega_H\} \in (0,1)^4$. If for both $e = \varepsilon$ and $e=H$ we have
\begin{equation}\label{eq:tech}
    \beta R \sum_{e' \in \{ \varepsilon,H  \} } P(e,e') \max \left\{ \frac{h(0,e)}{h(e - h(0,e) , e'  )},\frac{h(\bar{a},e)}{h(R\bar{a} +  e - h(\bar{a},e),e'   )} \right\}^\gamma \leq 1,
\end{equation} then $h$ is a subsolution for the \citet{coleman1990solving} operator \eqref{eq:Coleman}, and, consequently, $a'=g(a,e) \leq Ra + e - h(a,e)$ for all $(a,e)$. Moreover,
\begin{equation}\label{eq:UB2}
    A(R) \leq \bar{A}(R) \equiv  \mathbf{1}^\top   (I - R P^\top D_\eta )^{-1} P^\top D_\omega \tilde{e},
\end{equation}
where $D_\eta = \begin{pmatrix} 1 - \eta_\varepsilon & 0 \\ 0 & 1-\eta_H \end{pmatrix}$, 
$ D_\omega = \begin{pmatrix} 1 - \omega_\varepsilon & 0 \\ 0 & 1-\omega_H \end{pmatrix}$, and $\tilde{e} = (p_\varepsilon e_\varepsilon,p_H H)^\top$.
\end{prop}
The condition \eqref{eq:tech}, which is easy to check in practice, is sufficient for $h$ being a subsolution to $c$. $a=0$ and $a = \bar{a}$ show up because the numerators and denominators of the bracket terms are linear in $a$, implying that the fractions take their maximums at the endpoints. The condition looks complicated, but it sidesteps having to check the subsolution for all $a$, which would be impossible numerically (since $a$ is a continuous variable) and thus undermine the goal of an analytical example with proven multiplicity.

The key upper bound \eqref{eq:UB2} basically just integrates the individual saving upper bounds at the stationary distribution. It is convenient to derive it in matrix form because the transpose of the Markov transition matrix $P$ appropriately allocates total capital by income today to capital by income tomorrow.

\section{Multiplicity}\label{sec:mult}

The bounds I have established reveal a simple algorithm for trying to find \citet{Aiyagari1994} examples with multiple steady states when there are two individual income states: 
\begin{enumerate}
    \item Impose the parameter restriction $\left(\beta(1-\pi)\right)^\frac{1}{\gamma-1} > 1 - (1-\alpha)\delta $. By Theorem \ref{thm:verylow}, if we set $\varepsilon$ to a very small number there will be steady state at $r_{\alpha \delta} \approx \alpha \times \delta$, and we will be able to find a rental rate $r_L$ slightly greater than $r_{\alpha \delta}$ at which there is excess supply ($Z(1+r_L - \delta)>0$).
    \item For each $R \in (0,1)$, find parameters $\{\eta_\varepsilon,\omega_\varepsilon,\eta_H,\omega_H\} \in (0,1)^4$ that satisfy the technical condition \eqref{eq:tech} and calculate $\bar{A}(R)$.
    \item If $\bar{A}(R) < D(R)$ for some $r_H \in (r_L,\delta)$, then there must be at least three steady states because supply $A(R)$ will again cross demand $D(R)$ as $r$ goes from $r_L$ to $r_H$, and then, by \citet{Acikgoz2018}, $A(R)$ will rise above $D(R)$ and ultimately diverge as $r \rightarrow 1/\beta -1 + \delta$ (meaning one can always find $\bar{r} \in (\delta,1/\beta -1 + \delta)$ with $Z(1+\bar{r} -\delta)>0$.
\end{enumerate}

Here is a parameter set that must yield at least three steady states in the canonical \citet{Aiyagari1994} model with two individual income states:
\begin{align*}
    \beta &  = 0.30 \\
    \gamma & = 3 \\
    \delta & = 0.90 \\
    \alpha & = 0.30 \\
    H & = 1 \\
    \varepsilon & = 10^{-12} \\
    N^* & = 28 \\
    \lambda & = 0.0001 \\
    \pi & = 0.001 \\
    (\eta_\varepsilon,\omega_\varepsilon) &  =  (0.14,0.65)\\
    (\eta_H,\omega_H) & =  (0.33,0.73)  
\end{align*}
At these parameters, the condition for Theorem \ref{thm:verylow} is satisfied:
$$ \left(\beta(1-\pi)\right)^\frac{1}{\gamma-1} \approx 0.55  > 0.37  \approx 1 - (1-\alpha)\delta.       $$
Now consider the rental rates
$$  0 < (\underline{r} =0.2) < (\alpha\delta= 0.27) < (r_L =0.3) < (r_H=0.6) < (\delta = 0.9),   $$
at which we have
\begin{align*}
    Z(1+\underline{r}-\delta) \leq & \bar{\bar{A}}(1+\underline{r}-\delta) - D(1+\underline{r}-\delta) \approx  -0.65 < 0  \\
    Z(1+r_L-\delta) \geq & \underline{A}_{N^*,\varepsilon}(1+r_L-\delta) - D(1+r_L-\delta) \approx  0.05 > 0  \\
    Z(1+r_H-\delta) \leq & \bar{A}(1+r_H-\delta) - D(1+r_H-\delta) \approx  -0.19 < 0,
\end{align*}
and the LHS of \eqref{eq:tech} is around $0.96<1$ and $0.90<1$ for $e=\varepsilon$ and $e=H$, respectively. And we will always be able to find $\bar{r}\in (\delta,1/\beta -1 +\delta)$ with $Z(1+\bar{r} - \delta)>0$ since $D$ is strictly decreasing while $A$ diverges.

\begin{figure}
\begin{center}
    \includegraphics[width = 15cm]{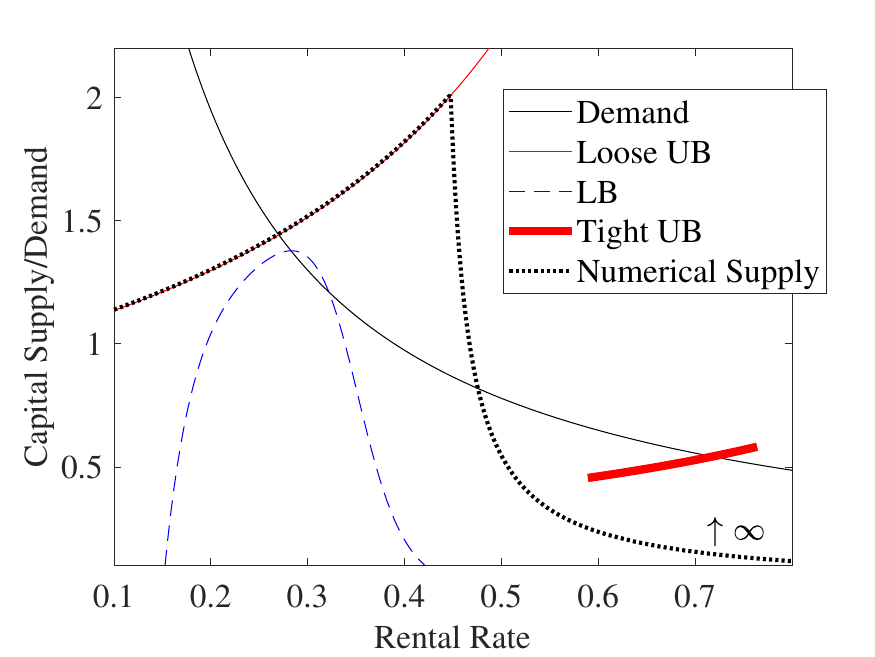}
    \caption{\textbf{Multiplicity Example}}
  \label{fig:ex}
  \end{center}
  \small
  Note: For the parameters given in Section \ref{sec:mult}, the figure plots capital demand $D$, the loose capital supply upper bound $\bar{\bar{A}}$, the lower bound on capital supply $\underline{A}$, and the tight supply upper bound $\bar{A}$. The black dotted line is the numerically-computed stationary capital supply $A$. Note that the tight upper bound $\bar{A}$ is plotted with fixed $(\eta,\omega)$'s, but only for rental rates at which \eqref{eq:tech} is satisfied. $\uparrow \infty$ indicates that we know supply will ultimately diverge as $r \rightarrow 1/\beta -1 +\delta$ (which is $>3$ in this example).
\end{figure}

So there must be at least three steady states, one near $\alpha \times \delta$, one between $r_L$ and $r_H$, and the last one for some higher $r$.\footnote{I have not ruled out that there could be more steady states, but \citet{WalshYoung2026} never found an example with more than three in their extensive search.} Figure \ref{fig:ex} plots the bounds in this example, alongside a numerical computation of stationary supply for this example.\footnote{For each $r$, I solve the individual agent problem via the endogenous grid method, simulate a panel of agents for many periods, and calculate $A$ as average capital in the last period.} Note that, consistent with Proposition \ref{lem:Alim}, numerical supply is very close to $\bar{\bar{A}}$ (which is \emph{not} imposed in the computation) for $R< \left(\beta(1-\pi)\right)^\frac{1}{\gamma-1}$, since $\varepsilon$ is so small. Numerically, there are three steady states at $r_{\alpha \delta} \approx 0.27$, $r_2\approx0.48$, and $r_3 \approx 3.22$.

\section{Conclusion}\label{sec:conclusion}

I have provided the first proof that the model of \citet{Aiyagari1994} can have multiple steady states. On the one hand, this is not surprising since multiplicity of some form or another can easily arise with heterogeneous agents. On the other hand, the literature provides few theoretical results on this type of model, and computational multiplicity examples were only recently provided, despite decades of investigation and empirical applications of this framework. Indeed, until now, the possibility of multiple steady states in the vanilla \citet{Aiyagari1994} model was an open question.\footnote{The computed examples of \citet{Acikgoz2018} and \citet{WalshYoung2026} strongly suggested multiplicity was possible, but they didn't prove it.} I have only covered the case with two individual income states, but there is no reason to believe my approach wouldn't extend to more states. In any case, including more than two states should make multiplicity more likely by introducing more degrees of freedom. As described in \citet{WalshYoung2026}, there are many potentially interesting applications of steady-state multiplicity in Bewley models, and I hope that my new bounds will inspire both analytic and computational ventures along these lines.



\clearpage
\newpage

\appendix

\section{Proofs}\label{app:proofs}

\subsection{Proof of Proposition \ref{lem:Alim}}

Choose $\nu>0$. Dividing the numerator and denominator of \eqref{eq:LB} by $\left( (\beta R)^Nq_N \right)^{1/\gamma}$, we can write the lower bound as
$$ \underline{A}_{N,\varepsilon}(R) = \frac{ L_\varepsilon - \frac{1-R^N}{1-R}  \frac{\varepsilon}{\left( (\beta R)^Nq_N \right)^{1/\gamma}}}{\frac{R^N}{\left( (\beta R)^Nq_N \right)^{1/\gamma}} + (1-R) },  $$
where 
$$ q_N = (1-\pi)^{N-1} \min\{ 1-\pi,\lambda     \}.  $$
Since 
\begin{align*}
   \frac{R^N}{\left( (\beta R)^Nq_N \right)^{1/\gamma}} = & \left(\frac{1-\pi}{\min\{ 1-\pi,\lambda     \}} \right)^{1/\gamma} \left( \frac{R^{N \gamma}}{ \beta^N R^N (1-\pi)^{N} } \right)^{1/\gamma} \\ 
   = & \left(\frac{1-\pi}{\min\{ 1-\pi,\lambda     \}} \right)^{1/\gamma} \left( \frac{R}{ (\beta (1-\pi))^{\frac{1}{\gamma-1}} } \right)^{N(\gamma-1)/\gamma},
\end{align*}
by $R <\left(\beta(1-\pi)\right)^\frac{1}{\gamma-1}$ we have
\begin{equation}\label{eq:LBstep1}
   \lim_{N\rightarrow \infty} \frac{R^N}{\left( (\beta R)^Nq_N \right)^{1/\gamma}} = 0. 
\end{equation}
Now define $$f(N) = \frac{\left( (\beta R)^Nq_N \right)^{1/\gamma}}{N}.$$ The numerator goes to zero as $N$ goes to infinity (by $\beta,R<1$), so $\lim_{N\rightarrow\infty} f(N) = 0$. Then using \eqref{eq:LBstep1} and replacing $\varepsilon$ with $f(N)$, it follows that
$$ \lim_{N\rightarrow \infty}  \underline{A}_{N,f(N)}(R)  = \frac{p_HH}{1-R},$$ where I used that $\lim_{\varepsilon \downarrow 0 } L_\varepsilon= \lim_{N \rightarrow \infty } L_{f(N)}= p_HH$ and $\lim_{N \rightarrow \infty} \frac{1-R^N}{1-R}(1/N) = 0$ by $R<1$.
But since we also have $\lim_{N\rightarrow \infty}\frac{L_{f(N^*)}}{1-R} = p_HH/(1-R) < \infty$, and, as I showed in the main text, $\underline{A}_{N^*,f(N^*)}(R) < \frac{L_{f(N^*)}}{1-R}$, then we can find $N^*>1$ such that
$$ \frac{L_{f(N^*)}}{1-R} - \underline{A}_{N^*,f(N^*)}(R) < \nu.    $$
Finally, set $\bar{\varepsilon} = f(N^*) $ and recall from the main text that  $\underline{A}_{N,\varepsilon}(R) \leq A(R) \leq \frac{L_{\varepsilon}}{1-R}$ for all $N>1$ and $\varepsilon>0$. The proposition then follows from noting that $\underline{A}_{N,\varepsilon}(R)$ is strictly increasing as $\varepsilon$ falls from $\bar{\varepsilon}$ towards zero for fixed $N$, and $L_{\varepsilon}$ strictly decreases to $p_HH$. $\blacksquare$.

\subsection{Proof of Theorem \ref{thm:verylow}}
Choose $\iota > 0$. Note first that at rental rate $r= \alpha \delta$ (gross return $R = 1 - (1-\alpha) \delta$), capital demand ($D(R) =  \frac{\alpha}{1-\alpha} \frac{L_\varepsilon}{R -1 + \delta}$) is equal to the loose upper bound on capital supply, as noted in the main text:
$$ D(1-(1-\alpha)\delta) = \frac{L_\varepsilon}{(1-\alpha)\delta } = \bar{\bar{A}}(1-(1-\alpha)\delta),$$ for any $\varepsilon>0$. 

\bigskip
\emph{Step 1:} $Z(1+r_L - \delta) > 0$

By the assumption of the theorem $\left(\beta(1-\pi)\right)^\frac{1}{\gamma-1} > 1 - (1-\alpha)\delta $, the intersection of capital demand and the loose supply upper bound at $r = \alpha \delta$ occurs in the $R$ range admissible in Proposition \ref{lem:Alim}. Now, since capital demand is strictly decreasing in $r$ and  $\frac{L_\varepsilon}{1- R } = \frac{L_\varepsilon}{\delta -r }$ is strictly increasing in $r$, we can find $r_L \in \left(\alpha \delta,\left(\beta(1-\pi)\right)^\frac{1}{\gamma-1} - 1 +\delta\right)$ such that $r_L-\alpha\delta<\iota$ and $\frac{L_\varepsilon}{1- R } > \frac{\alpha}{1-\alpha} \frac{L_\varepsilon}{R -1 + \delta}$. Importantly for the proof, we can take this $r_L$ independently of $\varepsilon$ because $L_\varepsilon$ is on both sides of $\frac{L_\varepsilon}{1- R } > \frac{\alpha}{1-\alpha} \frac{L_\varepsilon}{R -1 + \delta} = D(R)$.

Now, define $\nu>0$ by $$\frac{\nu}{2} = \frac{p_H H}{\delta-r_L } - \frac{\alpha}{1-\alpha} \frac{p_H H}{r_L}  >0,  $$
which is a lower bound on the distance between demand and the loose supply upper bound at $r_L$ since $L_\varepsilon \in (p_H H,H)$ as $\varepsilon$ varies between $0$ and $H$. By Proposition \ref{lem:Alim}, we can find $\bar{\varepsilon}>0$ and $N^*>1$ such that for all $\hat{\varepsilon} \in (0,\bar{\varepsilon})$ we have $\frac{L_{\hat{\varepsilon}}}{1-R} - \underline{A}_{\hat{\varepsilon},N^*}(1+r_L -\delta) < \frac{\nu}{2}$. Choose arbitrary $\varepsilon \in (0,\bar{\varepsilon})$. It follows that
\begin{align*}
    A(1 + r_L - \delta) - D(1 + r_L - \delta) = &  A(1 + r_L - \delta) - \frac{\alpha}{1-\alpha} \frac{L_\varepsilon}{r_L} \\
    \geq & \underline{A}_{\varepsilon,N^*}(1+r_L -\delta) - \frac{\alpha}{1-\alpha} \frac{L_\varepsilon}{r_L} \\
    >  & - \frac{\nu}{2} + \left(\frac{L_{\varepsilon}}{\delta-r_L}  -\frac{\alpha}{1-\alpha} \frac{L_\varepsilon}{r_L}\right) \\
    > & - \frac{\nu}{2} + \frac{\nu}{2} = 0,
\end{align*}
where the last inequality uses the definition of $\nu$ and $p_H H < L_\varepsilon$. So we have $r_L-\alpha\delta<\iota$ and $Z(1+r_L-\delta) > 0$, provided $\varepsilon \in (0,\bar{\varepsilon})$.

\bigskip
\emph{Step 2:} $Z(1+\underline{r} - \delta) < 0$

Recall that at $r = \alpha \delta$, the expression for firm demand is equal to the loose upper bound on capital supply $\bar{\bar{A}}(1+r-\delta) = \frac{L_\varepsilon}{\delta - r }$, as noted above, and recall that while firm demand is strictly decreasing in $r$, the upper bound on capital supply is strictly increasing. So we can find $\underline{r} \in (0,\alpha \delta)$ such that $\alpha \delta  - \underline{r} < \iota$ and $A(1+\underline{r} - \delta) \leq \bar{\bar{A}}(1+\underline{r}-\delta) < D(1+\underline{r} -\delta)$ (so $Z<0$). Note that, as with $r_L$, $\underline{r}$ can be chosen independently of $\varepsilon>0$ since $L_\varepsilon$ is on both sides of $\frac{L_\varepsilon}{1- R } < \frac{\alpha}{1-\alpha} \frac{L_\varepsilon}{R -1 + \delta} = D(R)$.

\bigskip
\emph{Step 3:} $Z(1+r_{\alpha \delta} - \delta) = 0$

We have already shown that $Z(1+\underline{r}-\delta)<0$ at some $\underline{r} \in (\alpha \delta - \iota,\alpha\delta) $. And we showed that for some $r_L \in (\alpha \delta,\alpha \delta + \iota)$ independent of $\varepsilon$, we have $Z(1+r_L-\delta)>0$ if $\varepsilon \in (0,\bar{\varepsilon})$. By \citet{Acikgoz2018}, $Z$ is continuous on $[\underline{r},r_L]$, so there exists $r_{\alpha \delta} \in (\underline{r},r_L)$ with $Z(1+r_{\alpha \delta} -\delta) = 0$ by the intermediate value theorem, and $r_{\alpha \delta}$ differs from $\alpha \times \delta$ by at most $\iota$ (which was arbitrary). $\blacksquare$

\subsection{Proof of Corollary \ref{cor:verylow}}
The corollary was an intermediate step in the Proof of Theorem \ref{thm:verylow}. $\blacksquare$

\subsection{Proof of Lemma \ref{lem:aub}}
Suppose the current individual state is $(a_t,e_t) \in [0,\bar{a}] \times \mathcal{E}$, where $\bar{a} = H/(1-R)$. By $c_t\geq0$, the budget constraint, and $e_t\leq H$, we have $$a_{t+1} = Ra_t + e_t - c_t \leq R\bar{a} + H = \frac{H}{1-R} = \bar{a}.$$$\blacksquare$

\subsection{Proof of Proposition \ref{prop:UB}}
The proof consists of two steps. First, I establish a subsolution for consumption and the upper bound on individual saving. Second, I aggregate the individual upper bounds in the stationary distribution to get an upper bound on aggregate stationary capital supply.

\bigskip
\emph{Step 1:} Consumption subsolution and upper bound on $a' = g(a,e)$

Suppose first that, for all $(a,e)$, we have
\begin{equation}\label{eq:tech2}
    \beta R \sum_{e' \in \{ \varepsilon,H  \} } P(e,e')  \left( \frac{h(a,e)}{h(Ra +  e - h(a,e),e'   )} \right)^\gamma \leq 1.
\end{equation}
Shortly, I will show that the condition \eqref{eq:tech} is sufficient for this. \eqref{eq:tech2} implies
\begin{align}
    h(a,e)^{-\gamma} \geq &  \beta R \sum_{e' \in \{ \varepsilon,H  \} } P(e,e')  \left( h(Ra +  e - h(a,e),e'   ) \right)^{-\gamma} \nonumber \\ & \implies \nonumber \\
     h(a,e)^{-\gamma} \geq & \max  \left\{  \beta R \sum_{e' \in \{ \varepsilon,H  \} } P(e,e')  \left( h(Ra +  e - h(a,e),e'   ) \right)^{-\gamma}  , (Ra + e)^{-\gamma}  \right\},   \label{eq:tech3} 
\end{align}
where the second inequality uses that $(\cdot)^{-\gamma}$ is strictly decreasing and $h(a,e)\leq Ra + e$ (by the assumed form of $h$).

Now compare \eqref{eq:tech3} with the implicit \citet{coleman1990solving} operator equation
\begin{equation}\label{eq:coleman2}
   (\mathcal{K}h(a,e))^{-\gamma} = \max  \left\{  \beta R \sum_{e' \in \{ \varepsilon,H  \} } P(e,e')  \left( h(Ra +  e - \mathcal{K}h(a,e),e'   ) \right)^{-\gamma}  , (Ra + e)^{-\gamma}  \right\}.
\end{equation}
Given a current consumption function $h$, this equation updates $h$ to $\mathcal{K} h$ by finding $\mathcal{K} h$ that solves \eqref{eq:coleman2} (pointwise). \citet{Acikgoz2018} shows that there is a unique solution. Because \eqref{eq:tech3} holds for all $(a,e)$, \eqref{eq:coleman2} implies that $h \leq \mathcal{K} h$. To see this, first note that the LHS of $\eqref{eq:coleman2}$ is strictly decreasing in $\mathcal{K} h(a,e)$. And the RHS is strictly increasing in $\mathcal{K} h(a,e)$ (because $h$ is strictly increasing in $a$ by assumption).

Evaluating $\mathcal{K} h(a,e)$ at $h(a,e)$ in \eqref{eq:coleman2}, by \eqref{eq:tech3} the LHS of \eqref{eq:coleman2} would be bigger than the RHS. Since the LHS is strictly decreasing and RHS is strictly increasing in $\mathcal{K} h(a,e)$, it follows that the true solution to the \citet{coleman1990solving} update must satisfy $\mathcal{K} h(a,e) \geq h(a,e)$. But $(a,e)$ was arbitrary, so $h \leq \mathcal{K}h$. Since the implicit \citet{coleman1990solving} operator is monotone and has a unique fixed point $c$ (by \citet{Acikgoz2018}), we can conclude
$$ h \leq  \mathcal{K}h \leq \mathcal{K}^2h \leq \dots \leq c,$$ so $h$ is a subsolution. The budget constraint then implies $a' = g(a,e) \leq Ra + e - h(a,e)$.

It remains to be shown that \eqref{eq:tech} from the statement of the proposition is sufficient for \eqref{eq:tech2} for all $a$. Since the numerator and denominator of $\frac{h(a,e)}{h(Ra +  e - h(a,e),e'   )}$ are linear in $a$ (by assumption), the maximum of $\frac{h(a,e)}{h(Ra +  e - h(a,e),e'   )}$ occurs at an endpoint, i.e. either $a = 0$ or $a = \bar{a}$, where $a$ can be taken to be bounded without loss of generality by Lemma \ref{lem:aub}. So
\begin{equation*}
    \beta R \sum_{e' \in \{ \varepsilon,H  \} } P(e,e') \max \left\{ \frac{h(0,e)}{h(e - h(0,e) , e'  )},\frac{h(\bar{a},e)}{h(R\bar{a} +  e - h(\bar{a},e),e'   )} \right\}^\gamma \leq 1
\end{equation*}
from the proposition statement implies 
\begin{equation*}
    \beta R \sum_{e' \in \{ \varepsilon,H  \} } P(e,e')  \left( \frac{h(a,e)}{h(Ra +  e - h(a,e),e'   )} \right)^\gamma \leq 1 \ \forall a,
\end{equation*}
which is \eqref{eq:tech2}, and we are done with \emph{Step 1}.

\bigskip
\emph{Step 2:} Upper bound on $A(R)$

The starting point is the individual supply upper bound $a' = g(a,e) \leq Ra + e - h(a,e)$, which, after substituting in the expression for $h$ becomes
$  g(a,e) \leq R(1-\eta_e)a + (1-\omega_e)e$. Integrating this at the stationary distribution (conditional on $e$) gives
\begin{equation}\label{eq:gint}
    A'_e \leq R(1-\eta_e) A_e + (1-\omega_e)p_e e,
\end{equation}
where $A'_e = \int_{s = e} g(a,s) d\Omega(a,s)$ is the supply of capital at $t+1$ from all agents who had income state $e$ at time $t$ (in the stationary distribution), and $A_e = \int_{s = e} a d\Omega(a,s)$ is the capital supply of agents with income $e$ in the stationary distribution. I also used $\int_{s = e} s d\Omega(a,s) = p_e e$. To ease notation, I suppress reference to the dependence of everything on $R$.

Now, let $\tilde{A}' = (A'_\varepsilon,A'_H)^\top$ denote the vector of $t+1$ capital supply from income states $\varepsilon$ and $H$ at time $t$, let $\tilde{A} = (A_\varepsilon,A_H)^\top$, and let $\tilde{e} = (p_\varepsilon e_\varepsilon,p_H H)^\top$. Clearly the total stationary capital supply is $A = \mathbf{1}^\top \tilde{A} = A_\varepsilon +A_H $, and total labor income is $L = \mathbf{1}^\top \tilde{e} $. Stacking, \eqref{eq:gint} in matrix form is
\begin{equation}\label{eq:amat}
    \tilde{A}' \leq R D_\eta \tilde{A} + D_\omega\tilde{e},
\end{equation}
where $D_\eta = \begin{pmatrix} 1 - \eta_\varepsilon & 0 \\ 0 & 1-\eta_H \end{pmatrix}$
and 
$ D_\omega = \begin{pmatrix} 1 - \omega_\varepsilon & 0 \\ 0 & 1-\omega_H \end{pmatrix}$.

Finally, from stationarity we have $\tilde{A} = P^\top \tilde{A}' $, where $P$ is the transition matrix for individual income. Combining this with \eqref{eq:amat} and rearranging, we get the upper bound
$$  A \leq \bar{A} =  \mathbf{1}^\top   (I - R P^\top D_\eta )^{-1} P^\top D_\omega \tilde{e},    $$
as we set out to prove. Note that
$$  (I - R P^\top D_\eta )^{-1} = I + R P^\top D_\eta + (R P^\top D_\eta)^2 \dots     $$
Since $P^\top D_\eta = \begin{pmatrix} (1-\pi)(1 - \eta_\varepsilon) & \lambda(1 - \eta_H) \\ \pi(1 - \eta_\varepsilon) & (1-\lambda)(1 - \eta_H) \end{pmatrix}$, all of the elements of $P^\top D_\eta$ are strictly between $0$ and $1$. Moreover, since the column sum of $P^\top D_\eta$ is $(1-\eta_\varepsilon,1-\eta_H)$ and $\eta_s \in (0,1)$, the spectral radius of $P^\top D_\eta$ is strictly is less than one. Therefore, since $R \in (0,1)$, the inverse exists and it is positive (because all of the terms in the series are positive). $\blacksquare$

\clearpage
\newpage

\printbibliography

\end{document}